\documentclass[10pt]{article}

\usepackage{fullpage}
\usepackage[utf8]{inputenc}
\usepackage[english]{babel}

\usepackage{setspace}
\usepackage{parskip}
\usepackage{titlesec}
\titlespacing{\section}{0pt}{*3}{*1}
\titlespacing{\subsection}{0pt}{*2}{*0.5}
\titlespacing{\subsubsection}{0pt}{*1.5}{0pt}

\usepackage[section]{placeins}
\usepackage{xcolor}
\usepackage{lineno}
\usepackage{hyphenat}

\PassOptionsToPackage{hyphens}{url}
\usepackage[colorlinks=true, linkcolor=blue, urlcolor=blue, citecolor=blue]{hyperref}

\usepackage[numbers,sort&compress]{natbib}
\usepackage{authblk}

\usepackage{graphicx}
\usepackage{longtable}
\usepackage{tabulary}
\usepackage{booktabs,array,multirow}
\usepackage{amsmath,amssymb,amsfonts}
\usepackage{siunitx}
\usepackage{textcomp}

\renewenvironment{abstract}
  {{\bfseries\noindent{\abstractname}\par\nobreak}\footnotesize}
  {\bigskip}

\begin{document}

\title{{Mesoscale flows in active baths} dictate the dynamics of semi-flexible
filaments}




\author[1]{Bipul Biswas\textsuperscript{\textdagger}}
\author[1]{Devadyouti Das\textsuperscript{\textdagger}}
\author[1]{Manasa Kandula\textsuperscript{\textasteriskcentered}}
\author[1]{Shuang Zhou\textsuperscript{\textasteriskcentered}}

\affil[1]{Department of Physics, University of Massachusetts Amherst}
\affil[ ]{\textsuperscript{\textdagger}These authors contributed equally.}
\affil[ ]{\textsuperscript{\textasteriskcentered}Corresponding authors: Manasa Kandula (\texttt{hkandula@umass.edu}), Shuang Zhou (\texttt{zhou@physics.umass.edu}).}

\vspace{-1em}

\date{}

\begingroup
\let\center\flushleft
\let\endcenter\endflushleft
\maketitle
\endgroup

\onehalfspacing

\selectlanguage{english}
\begin{abstract}
Semi-flexible filaments in living systems are constantly driven by
active forces that often organize into mesoscale coherent flows.
Although theory and simulations predict rich filament dynamics,
experimental studies of passive filaments in collective active baths
remain scarce. Here we present an experimental study on passive
colloidal filaments confined to the air--liquid interface beneath a
free-standing, quasi-two-dimensional bacterial film featuring
jet-like mesoscale flows. By varying filament contour length and
bacterial activity, we demonstrate that filament dynamics are governed
by its length relative to the characteristic size of the bath. Filaments
shorter than the jet width exhibit greatly enhanced translation and
rotation with minimal deformation, while long filaments show dramatic
deformation but less enhanced transport. We explain our findings through
the competition between the active viscous drag of the bath and passive
elastic resistance of the filaments, using a modified elastoviscous
number that considers the mesoscale flows.~%
\end{abstract}%

\sloppy

\textbf{Keywords:}~

\begin{quote}
filament propulsion, active bacterial bath, collective motion,
elastoviscous number, softening, conformational change,~ mesoscale
flows, emergent dynamics
\end{quote}

\par\null

\par\null

\section*{Introduction}

{\label{727603}}

Semi-flexible filaments are ubiquitous in living systems, ranging
from actin\citep{pollard_actin_2009,blanchoin_actin_2014} and microtubules\cite{suzuki_spatial_2017} that
sculpt cell architecture, to DNA and chromatin fibers that organize
genetic material ~\citep{zidovska_rich_2020}. Unlike passive filaments in
equilibrium baths, whose dynamics are driven only by thermal
fluctuations, biological filaments are constantly driven by active
forces that often self-organize into mesoscale motions coherent over
distances much larger than individual active agents, examples including
chromatin flows, metabolic eddies, motor-driven streaming, and vortices
around organelles\citep{kimura_endoplasmic-reticulum-mediated_2017,foffano_dynamics_2016}.~ As a result, biofilaments often
exhibit rich and non-equilibrium transport and conformational features
that are critical for their biological functions in the crowded, active
environments. Understanding how they emerge from the interplay between
filament features, such as their length, flexibility, and chemical
composition, and activity features, such as its size, strength,
duration, distribution, and directionality, provides profound insights
into the critical biological processes\citep{weady_conformations_2024}. ~

Motivated by the biological relevance, a substantial body of
theoretical and computational work has explored the dynamics of
filaments influenced by incoherent active forcing.~Simulations of
passive filaments in dilute active baths predict a wealth of filament
dynamics, including buckling, anomalous stiffening,~ regimes of negative
dissipation~\citep{kikuchi_buckling_2009}, anomalous swelling~ for flexible
filaments\citep{eisenstecken_conformational_2016,vandebroek_dynamics_2015} , and contracting-swelling transition for
semiflexible filaments\citep{eisenstecken_conformational_2016}. Alternatively, by embedding
activity through self-propelled or force-generating monomers, active
filament simulations predict that even weak internal driving leads to
translational and rotational propulsion and activity-induced
instabilities driven by spontaneous symmetry breaking via self-generated
hydrodynamic flows~\citep{jayaraman_autonomous_2012}, and large conformational changes,
including a self-amplifying coil--stretch transition and strongly
elongated filament conformations\cite{mahajan_self-induced_2022} at high activity
level.~

Experimental studies have primarily focused on active
filaments~\citep{biswas_emergent_2025,kumar_emergent_2024,nishiguchi_flagellar_2018,haque_propulsion_2023,yang_reconfigurable_2020,biswas_linking_2017}, with only a few investigations addressing
passive filaments in dilute active baths, and even fewer exploring
active baths exhibiting collective states. In studies of passive
filaments in~structureless, dilute active baths, hairpin-like states and
propulsion of semi-flexible~filaments were observed in granular
experiments where thermal fluctuations are absent ~\citep{anderson_polymer-chain_2022,aporvari_anisotropic_2020}.
Flexible passive filaments show localized bending in addition to
propulsion when interacting with individual bacteria
~\citep{zhang_configurational_2023}, and show coiling and anisotropic diffusion when
interaction with a collective bacteria bath ~\citep{liu_enhanced_2025} . But
the work does not explicitly address the role of collective flows of the
bath~ and the intrinsic rigidity of the filament. On the other hand,
insights from studies of fibers in classical turbulent flows have
demonstrated a rich spectrum of dynamics, including buckling, twisting,
sustained spinning, tumbling, and alignment with
vortices\citep{allende_stretching_2018,gay_characterisation_2018,oehmke_spinning_2021,olivieri_fully_2022}.~ These results highlight the central role
of flow structure and its characteristic length scales in governing
filament deformation and transport, pointing to unexplored regimes of
filament dynamics in coherent active flows.~~

To bridge this gap, we create an experimental system that provides
simultaneous and independent control over a broad spectrum of filament
properties and active environments, while allowing easy quantification
of both filament and bath dynamics through microscopy imaging. The
system is composed of a free-standing, quasi-two-dimensional bacterial
film~\citep{sokolov_concentration_2007}~ with colloidal filaments dwelling at the
bottom air-liquid interface due to gravity. By varying the contour
length of the filaments and the activity of the bacterial bath
independently, we show that the filament length, relative to the
characteristic length scales of the ``bacterial jets'' (elongated
coherent flows emergent in high-activity baths), dictates their
transport and conformational dynamics.~While short filaments show strong
translation and rotation with little deformation, long filaments show
strong deformation and relatively less enhanced transport. We explain
our findings through the dominant jet-filament one-way interactions that
bend the filaments only when they are longer than the jet width. We
quantify filament deformation using the average curvature and explain
its dependence on activity through the competition between the active
viscous drag and passive elastic resistance, using a modified,
dimensionless elastoviscous number.~

\section*{Results}

{\label{167492}}

\subsection*{Enhanced Transport and Softening of Filaments Due to
Bacterial
Activity}

{\label{293753}}

The transport behavior of the colloidal filaments are distinct in active and passive baths. To quantify their translational motion, we track the centers of mass(CM) of the filaments and compute their mean square displacements(MSD) as a function of time, $\langle |\mathbf{r}(t+\tau)-\mathbf{r}(t)|^2 \rangle \propto t^{\alpha}$. In passive baths (suspensions of heat-killed bacteria), the CM motion remains diffusive and sub-diffusive ($\alpha \leq 1$) over the experimental time scale $\sim 75$ seconds (SI). In contrast, in active baths, the motion is ballistic for short times ($\alpha=2$ for $0<t < 0.5s $) and becomes diffusive ($\alpha \sim 1 $)at longer times (Fig 1c). Further, the average instantaneous velocities of the filament CM, $\langle V_{CM}\rangle$ scale roughly linearly with the average velocity of the bacterial bath, $\langle |\mathbf{v}|\rangle$, (SI). For the active cases, the MSD for the filament ends, $r_{end}$, shows an even stronger increase than that of $r_{CM} $ for all times(SI). To quantify the rotational motion, we track the orientation of the end-to-end vector of a filament and examine the corresponding mean-squared angular displacement, $\langle |\theta (t+\tau)-\theta(t)|^2 \rangle\propto t^{\alpha}$. Similar to the CM displacement, the $\langle |\theta (t+\tau)-\theta(t)|^2 \rangle$ is ballistic or super-diffusive at short times and transitions into diffusive motion at a later time (Fig 1d).

In addition to being transported by the active flows, the filaments undergo deformations and conformational changes (Fig 2 snapshots)depending on $\langle |\mathbf{v}|\rangle$. We characterize the instantaneous filament conformations in a configuration space spanned by the normalized radius of gyration $\widetilde{R}_g=R_g/L_C$ and the acylindricity $A^2$, both calculated from the gyration tensor $ \mathbf{G} $ (Materials and Methods). Here, $R_g=\sqrt{\lambda_1+\lambda_2}=\sqrt{Tr(\mathbf{G})}$ is the radius of gyration, which quantifies the spatial extent of the filament configuration, $L_C$ is the contour length of the filament, $A^2=\frac{\lambda_2-\lambda_1}{R_g^2}$ measures the shape asymmetry, and $\lambda_{1}$ , $\lambda_{2}$ are the two eigenvalues of $\mathbf{G}$. For example, $(A^2, \widetilde{R}_g )=\left(1, \,\frac{1}{2\sqrt{3}}\right)$ and $\left(0,\,\frac{1}{2\pi}\right)$ correspond to a perfectly straight and circular filament, respectively\citep{anderson_polymer-chain_2022}. With increasing activity while short filaments remain straight and occupy the same corner in the configuration space (Fig 2a), longer filaments expand to a larger area, exploring states of more compact configurations represented by smaller values of $\widetilde{R}_g$ and $A^2$ (Fig 2b,c). Remarkably, for very long filaments $L_c>200\mu m$, the increased spread of $A^2$ at low $\widetilde{R}_g$ values indicate the presence of folded-over hairpin-like structures (Fig 2c snapshots). The "softening" effect of the filaments is also reflected in the the rapid decay of the averaged orientational correlations of the tangent vectors, $\hat{t}$, of the filament centerline,  $f(s)=\langle \hat{t}(s\prime + s) \cdot \hat{t}(s\prime) \rangle_{s\prime} $ , as a function of the separation distance $s$ between two segments along the filament(SI).

\par\null

\subsection*{Mesoscale Structures of the Active
Baths}

{\label{711815}}

The dynamics of the passive filaments observed here arise directly from the emergent coherent structures of the active bacterial bath. The synthetic colloidal filaments used are rigid (bending rigidity $B \approx 2\times 10^{-21} \mathrm{J m}$)\citep{yang_superparamagnetic_2018} and much larger than the bacteria ($ d=4.2\mathrm{\mu m}$, $L_c=(15-600)\mathrm{\mu m}$). Therefore, in an passive bath without active bacteria (SI), the thermal agitation causes only a negligible motion and deformation of passive filaments. Even in dilute suspensions of active bacteria the filaments remain largely static (SI). Although pusher-type bacteria, such as \textit{Bacillus subtilis} used in this work, are known to preferentially swim near interfaces \citep{berke_hydrodynamic_2008,lemelle_counterclockwise_2010} and can collide directly with the filaments, the relatively large film thickness ($h \approx 50\,\mu\mathrm{m}$) allows bacteria to rapidly slide past rather than exert sustained forces. As a result, individual bacterium--filament encounters do not produce appreciable filament deformation or transport. The combination of large filament size, high rigidity, and weak quasi-two-dimensional confinement therefore renders the dynamics qualitatively different from prior studies in the dilute limit of active agents, where persistent particle--filament contacts give rise to anomalous swelling and hairpin formation\citep{vandebroek_dynamics_2015,kikuchi_buckling_2009, zhang_configurational_2023,harder_activity-induced_2014}.

At high bacterial concentrations, bacterial motion self-organizes into collective flows, characterized by active turbulence consistent with previous experimental and theoretical studies \citep{dombrowski_self-concentration_2004, wensink_meso-scale_2012}. The activity level of the bath can be characterized by its mean speed, $\langle|\mathbf{v}|\rangle = 10$--$60\,\mu\mathrm{m/s}$. For baths with $\langle|\mathbf{v}|\rangle \gtrsim 10\,\mu\mathrm{m/s}$, pronounced coherent structures emerge as elongated jets (Fig.~3a), whose characteristic velocities significantly exceed those of vortices (SI). While the orientations of the jets remain isotropically distributed, their velocity increases with $\langle|\mathbf{v}|\rangle$ (SI). Crucially, movement of filament segments are highly correlated with the local flow velocity (SI), showing that filament dynamics and deformations are dictated by these jets. Additionally, jets velocity is unchanged after encountering the filaments (SI), indicating that the one-way hydrodynamic coupling allows the filaments to respond to the collective bath without imposing feedback on its dynamics.

To quantify the average size of the jets, we compute the anisotropic spatial velocity-velocity correlations parallel and perpendicular to the local velocity direction (Materials and Methods), from which we extract the characteristic jet length $l$ and width $w$. Across the full activity range explored, the jet width remains approximately constant, $w = 25.5 \pm 10\,\mu\mathrm{m}$ (Fig.~3c, SI), without a clear dependence on $|\mathbf{v}|$. This observation is consistent with earlier studies of bacterial turbulence, which report that the characteristic wavenumber $\langle |q| \rangle$ is largely insensitive to activity level and is instead set primarily by the dimensionality of the system \citep{dunkel_fluid_2013}. $w$ values also align well with the width of strong flows ($w\prime$) directly observed in the PIV map of the bath, Fig 3a,b (SI).

\section*{Discussion~}

\subsection*{Length Dependence of Filament
Transport}

{\label{768696}}

In the active bath, filaments of all lengths exhibit ballistic behavior
at short times. However, the instantaneous translational and rotational
speeds are significantly higher for shorter filaments than for longer
ones (SI), resulting in correspondingly larger mean-squared
displacements. We attribute this length dependence to two primary
factors. First, longer filaments typically intersect multiple jets
simultaneously, whose velocities are uncorrelated. The net force and
torque transmitted to the filament are therefore partially averaged out,
leading to reduced overall transport.~Second, because the filaments are
semi-flexible rather than rigid, active flows acting on long filaments
preferentially induce local bending and deformation instead of
translating or rotating the entire filament.\textbf{~} In contrast,
filaments shorter than the jet width couple coherently to a single jet,
and hence get advected rapidly, or~undergo fast rotation in response to
local shear as a whole. As a result, when compared to long filaments,
shorter ones experience both higher instantaneous translational and
rotational speeds and a stronger dependence on activity level (Fig 1c
and d). For the same reason, the end segments of long filaments show a
higher instantaneous velocity and MSD than the entire filaments (SI). At
longer times, the translational and rotational dynamics of both short
and long filaments cross over to diffusive behavior because of the
finite lifetime of the jets and the lack of long-range temporal and
spatial correlations in the active flows. Similar crossover behavior has
been observed in other passive--active mixtures~\citep{wu_particle_2000}. ~

\subsection*{Length Dependence of Filament
Curvature}

{\label{694982}}

To probe the influence of coherent flow structures on filament conformation, we compute the average filament curvature, $\langle \kappa \rangle$, as a function of $\langle |\mathbf{v}|\rangle $. We find a clear length-dependent response. Filaments shorter than a threshold length ($L_c<L_{\mathrm{th}} \approx 25\,\mu\mathrm{m}$) behave effectively as rigid rods: their average curvature remains low and constant across activity levels (Fig.~4a). In contrast, for filaments with $L_c > L_{\mathrm{th}}$, $\langle \kappa \rangle$ increases monotonically with activity. Notably, the threshold length $L_{\mathrm{th}}$ coincides with the average jet width $\langle w \rangle$ of the active bath, Fig 3d. This correspondence suggests that filament deformation becomes significant only when a filament can couple to a jet over a sufficiently extended contour length. At high activity, the average curvature reaches $\langle \kappa \rangle \approx 0.012\,\mu\mathrm{m}^{-1}$, a value substantially smaller than the geometric estimate $1/\langle w \rangle \approx 0.04\mu\mathrm{m}^{-1}$ expected for a fully flexible filament confined within a jet of width $\langle w \rangle $. This reduction reflects both the finite bending rigidity of the filaments and the sparse, intermittent nature of the jets, which deform filaments only locally and transiently rather than imposing a uniform curvature along their entire length. We note that cases reported here only correspond to dynamic conformations responding instantaneously to newly encountered jets. Rare, dynamically arrested cases, such as long filaments with $L_c > 200 \mathrm{\mu m}$ exhibiting a fixed  configuration with high $\langle \kappa \rangle$ for extended duration (SI) are excluded. Inspection of the filament images and flow field reveals that such cases correspond to persistent U-shaped and hairpin-like conformations (SI). The formation of such shapes is typically initiated when a strong jet impinges near the center of a filament to induce a pronounced bend. Subsequent jets then advect the two arms closer together, completing the folding process. Once such shapes are formed, the narrow channel between the nearly parallel arms inhibits the penetration of new jets capable of unfolding the filament, thereby stabilizing the hairpin configuration and sustaining a high-curvature state over long times. Similar cases were reported in prior simulation work on filament immersed in dilute active particle, where the imbalanced number of active particles colliding the inside versus outside of the hairpin loop lead to a persistent conformation\citep{harder_activity-induced_2014}.

\subsection*{Competition between Viscous Forcing and Elastic Resistance
~}

{\label{ev}}

From the discussions above, we have demonstrated that thermal agitation or single-bacterium collisions are not strong enough, and the observed filament deformations are driven by collective flows, which must be resisted by the intrinsic filament rigidity. To characterize this competition, we compare the ratio of viscous forcing of the bath over the elastic resisting of the filaments, using the dimentionless elastoviscous number $\tilde{\mu}$. Conventionally, $\tilde{\mu}=\frac{8 \pi \eta \dot\gamma L^4}{B}$ consider the length of filaments $L$, together with flow properties such as dynamic viscosity of the fluid $\eta$ and shear rate $\dot{\gamma}$. It has been successfully implemented to explain filament dynamics in various systems, such as fiber shapes in inhomogeneous flow \citep{quennouz_transport_2015,liu_morphological_2018}, spermatozoa flagellum motion\citep{machin_wave_1958, gadelha_nonlinear_2010}, and actin filament conformations in active baths\citep{zhang_configurational_2023}. In a coherent active bath, however, the characteristic length scales of the bath, in our case the average jet width $w$, must enter the equation for the following reasoning. A single jet of width $w$ locally induces a shear flow with shear rate $\dot\gamma \propto |\Delta \mathbf{v}|/w$, where $|\Delta \mathbf{v}|=|\mathbf{v}_{jet}-\mathbf{v}_{bg}|$ is the typical net jet velocity relative to the background flow. When a section of a long filament encounters such a jet, the total force received is $\propto \eta \dot \gamma w^2$, and total torque $\propto \eta \dot \gamma w^3$. Since a long chain will have more chance to encounter jets, the total "forcing" on a chain is $\propto \eta \dot \gamma w^3 L_c$. Thus, for long filament $L_c>w$, the elastoviscous number is modified as $\tilde{\mu}_l=  \frac{8 \pi \eta \dot\gamma w^3 L_c}{B}$. For a short filament $L_c<w$, the equation remains the original form of $\tilde{\mu}_s= \frac{8 \pi\eta \dot\gamma L_c^4}{B}$, since the role played by jet width is replaced by filament length.

Following this argument, we plot $\langle \kappa \rangle$ vs $\tilde{\mu}_{s,l}$ (Fig 4b). We use $\eta \approx 1 \mathrm{mPa s}$ as the passive viscosity of the bacterial suspension with a volume fraction $\phi = 5 - 10 \%$, $\dot \gamma=(0.1-2) \mathrm{s}^{-1}$ the average shear rate measured from PIV of the bath, and $B = 2\times 10^{-21} \mathrm{J m}$ the bending rigidity of the colloidal filaments. Here, $\eta$ and $B$ are system constants, and hence the observed trend will not depend on them; $\dot \gamma$ and $w$ take into account the activity of the bath and statistically averaged feature sizes. We find that $\langle \kappa \rangle$ of different filament lengths show a unified threshold behavior on $\tilde{\mu}$, with a transition occurring close to $\tilde{\mu} =1 $. It is interesting to compare our results with the results of semi-flexible fiber in homogeneous shear flow \citep{liu_morphological_2018}, where a similar transition from tumbling to C-shape deformation is observed, but at $\tilde{\mu} \approx 10^3$. In their case, the Euler buckling due to compressive stress causes the formation of C-shape and subsequent U- and S- shapes, therefore requiring a much higher shear rate. However, in our case, since the jets provide transverse forces to directly bend the filament, the deformation occurs at a much lower $\tilde{\mu}$ values. In another word, the "softening" effect of transverse coherent flows in the active baths are orders of magnitude more effective than featureless shear flows.

In summary, the transport and conformation dynamics of semi-flexible
colloidal filaments in the coherent bacterial baths show that mesoscale
flows differentiate filaments by lengths and impart them with distinct
behaviors through a competition between active viscous drag and passive
elastic resistance. The transvers nature of the coherent jets bend long
filaments more effectively, evidenced by a transitional elastoviscous
number near unity.~ ~

In this study, we focused on a specific combination of filament and
active bath. Future projects could expand the exploration by altering
filament rigidity, length, and bath structure to further verify the
effectiveness of the modified elastoviscous number. Additionally, our
study are limited to 2D dynamics of filament receiving one-way
influences from the bath. In living systems, filament dynamics are often
3D and can provide substantial feedback to the active environments,
influencing its dynamics and energy flow. Experimental studies on these
more complex scenarios can yield novel insights for living systems and
stimulate future theoretical and simulation studies.~ ~

\section*{Materials and Methods}

{\label{806609}}

\subsection*{Bacterial Culture}

{\label{953715}}

\textit{Bacillus subtilis} cells of strain 1085 were streaked onto LB agar plates from -80\textdegree C stabilates, and incubated overnight at 33\textdegree C. Colonies from the agar plates were used to inoculate Terrific Broth(Sigma Aldrich) medium in 25ml cell culture flasks with vented caps for 6-7 hours at 33\textdegree C until early stationary stage with a plateaued concentration of $c_0 \sim 8\times10^8\,\mathrm{mL}^{-1}$. After incubation the cells were washed twice with de-ionized water by repeated centrifugation and resuspension to remove byproduct produced in bacterial metoablism and multiplication. Finally the suspension was centrifuged  at $\approx 12,000 g$ for 5 mins and concentrated to $50C_0$. During the experiments, the bacteria remain suspended in water instead of growth media to minimize the bio-production that can potentially change the viscoelasticity of the environment. During the experimental period, we do not observe velocity change, manifesting a constant activity level throughout the experiments.

\subsection*{Colloidal Filaments Fabrication}

{\label{colloids}}

 To fabricate filaments, we adapted the recipe outlined in prior reports \citep{biswas_emergent_2025, yang_superparamagnetic_2018}. We use carboxylic acid (-COOH) functionalized polystyrene (PS-COOH) colloids of $4.5 \mathrm{\mu m}$ diameter with a super-paramagnetic core (ProMag\textsuperscript{\tiny\textregistered} Microspheres, Polysciences). The particles are surface functionalized by overnight incubation at room temperature in $6\mathrm{mg/mL}$ aqueous solution of $1,4$-dithiothreitol (DTT, Sigma-Aldrich) and $2\mathrm{\mu L/m L}$ of $1,8$-diazabicycloundec-7-ene (DBU, Sigma-Aldrich). The functionalized colloids are cleaned with deionized (DI) water. Subsequently, $\approx 10\mu l$ of the suspension is added to  $\approx 60\mathrm{\mu L}$ of $5 \mathrm{m g/m L}$ aqueous solution of $40kDa$ 4-armed maleimide-functionalized polyethylene glycol (JenKem Technology). The vial was then placed in a magnetic solenoid coil and a DC electric current of about $3.5-3.7 A$ was applied, which produces a unidirectional DC magnetic field of $ 25-30\mathrm{mT}$. We carry out this reaction for about $25$ minutes. An electric fan is used to avoid overheating of the coil and to maintain the temperature at $\approx$ 60\textdegree C.

\subsection*{Experimental Setup}

{\label{setup}}

After mixing the colloidal filaments into the dense bacterial suspension, we deposited approximately $6\mathrm{\mu L}$ 
of the mixture onto two pairs of crossed glass fibers mounted on a custom-built sliding stage. By separating the two crossed fiber pairs and increasing the area of the parallelogram, we precisely reduce the thickness of the resulting bacterial film close to 100$\mu m$. This design was adapted from \citep{sokolov_concentration_2007}. The stage was placed inside an environmental control chamber, allowing independent regulation of humidity and oxygen concentration. In addition, we used deep-blue illumination(SI) to further modulate bacterial activity, providing an additional independent control.

\subsection*{Gyration tensor to quantify filament
conformations}

{\label{gyration}}

The normalized radius of gyration $\widetilde{R}_g=R_g/L_C$ and the acylindricity $A^2$ are derived from the two-dimensional gyration tensor, $G_{x y} = 1/N\sum_{i=1}^{N} (x_i-x_{cm}) (y_i-y_{cm})$, where $N$ is the number of points, $(x_i,y_i)$ coordinates of the centroid of the $i$th pixel along the ordered center line of the filament and $(x_{cm},y_{cm})$ the center of mass of the filament in any given conformation. Using the real eigenvalues, $(\lambda_1,  \lambda_2)$ of $G_{x y}$, we compute $R_g^2 = \lambda_1 + \lambda_2$, and $A^2 = (\lambda_2 - \lambda_1)/R_g^2$.

\subsection*{Estimation of Jet
Properties}

{\label{jet}}

To estimate the typical width of jets (regions of coherently moving bacteria at high velocity), we compute directional velocity--velocity correlations from the spatially-interpolated PIV velocity fields:

\begin{align*}
C_{vv}^{\parallel}(R)
&= \frac{\left\langle \mathbf{v}(t,\mathbf{r}) \cdot \mathbf{v}\!\left(t,\mathbf{r}+R\,\hat{\mathbf{e}}_{\parallel}(t,\mathbf{r})\right) \right\rangle_{t,\mathbf{r}}}{\left\langle \mathbf{v}^2\right\rangle},\\
C_{vv}^{\perp}(R)
&= \frac{\left\langle \mathbf{v}(t,\mathbf{r}) \cdot \mathbf{v}\!\left(t,\mathbf{r}+R\,\hat{\mathbf{e}}_{\perp}(t,\mathbf{r})\right) \right\rangle_{t,\mathbf{r}}}{\left\langle \mathbf{v}^2\right\rangle},
\end{align*}

where
\[
\hat{\mathbf{e}}_{\parallel}(t,\mathbf{r})
=\frac{\mathbf{v}(t,\mathbf{r})}{\| \mathbf{v}(t,\mathbf{r}) \|},
\qquad
\hat{\mathbf{e}}_{\perp}(t,\mathbf{r})
=\begin{pmatrix}0&-1\\ 1&0\end{pmatrix}\hat{\mathbf{e}}_{\parallel}(t,\mathbf{r})
\]
(in 2D), and the average is taken over all \((t,\mathbf{r})\) for which the displaced point lies within the valid interpolation domain.

For all cases, $C_{vv}^{\perp}(R)$ decays faster than $C_{vv}^{\parallel}$, showing that the "jets" are elongated structures with aspect ratio $ > 1$. We define the typical width, $w$, of the jets as the value of $R$ at which $C_{vv}^{\perp}(R=w)\approx 0.6 $.

\subsection*{}

{\label{191076}}

\section*{Acknowledgements}

D.D. and S.Z. acknowledge the support of NSF DMR- 2239551. B. B. and M.K.
acknowledge University of Massachusetts Amherst for the start-up funds.

\selectlanguage{english}
\newpage
\begin{figure}[h!]
\begin{center}
\includegraphics[width=0.70\columnwidth]{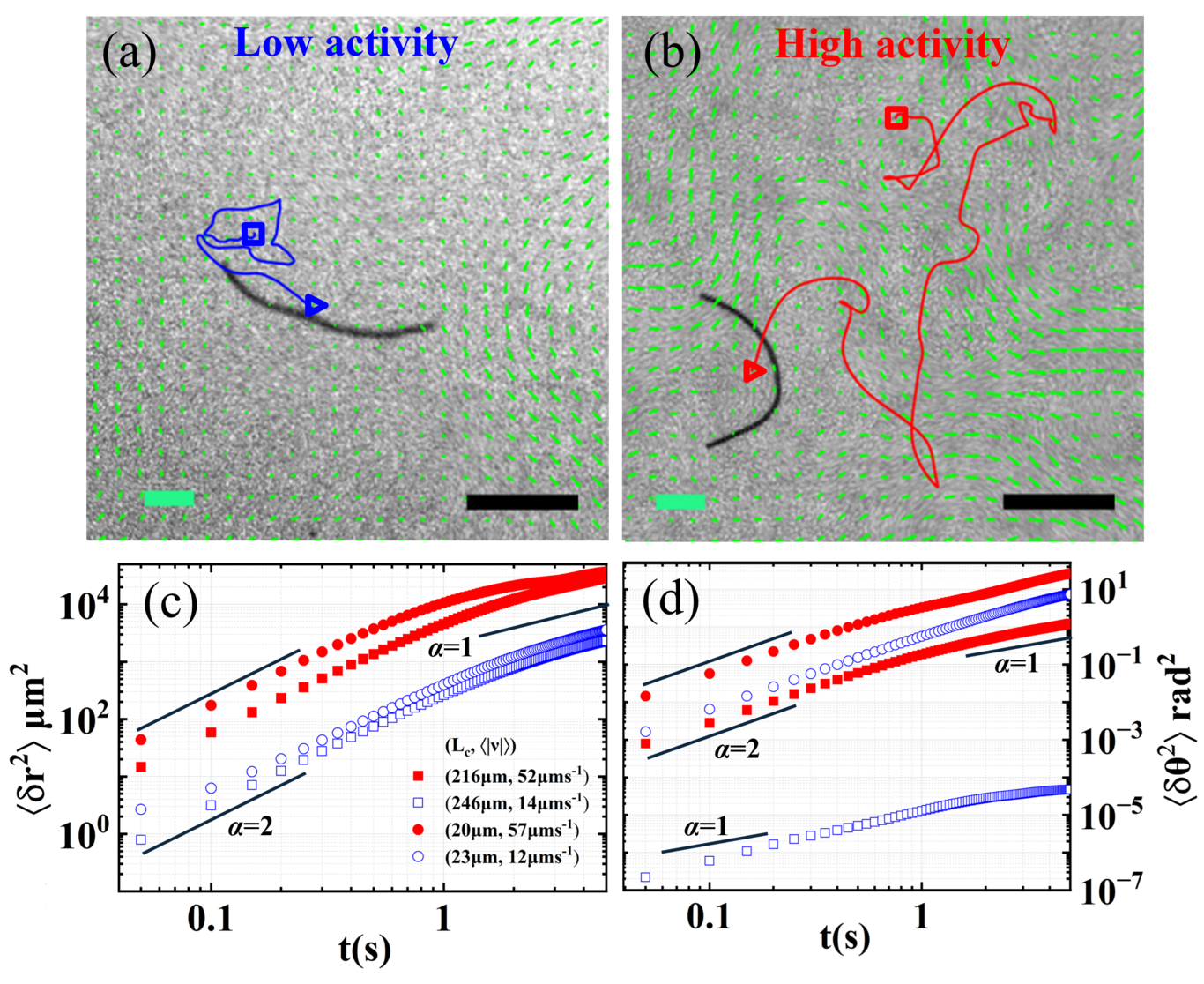}
\caption{{Transport dynamics of semi-flexible filaments in collective bacterial
baths. Typical trajectories of the filament CM under (a) low and (b)
high bacterial activity. Triangle and square mark the starting and
ending location of the filament during~10s of motion. Green arrows
represent the 2D flow field measured from PIV. (c) MSD of the filament
CM and (d) of the end-to-end vector for short~\(\left(l_c\sim20\mathrm{\mu m}\right)\) and
long~\((l_c\sim200\mathrm{\mu m})\) filaments under low\((|\mathbf{v}|\sim13\mathrm{\mu m/s})\) and high
~\((|\mathbf{v}|\sim 55\mathrm{\mu m/s})\) activities. Green scale bar:~\(100\mathrm{\mu m/s}\);
black scale bar:\(\mathrm{100\mu m}\)
{\label{543497}}%
}}
\end{center}
\end{figure}\selectlanguage{english}
\begin{figure}[h!]
\begin{center}
\includegraphics[width=0.70\columnwidth]{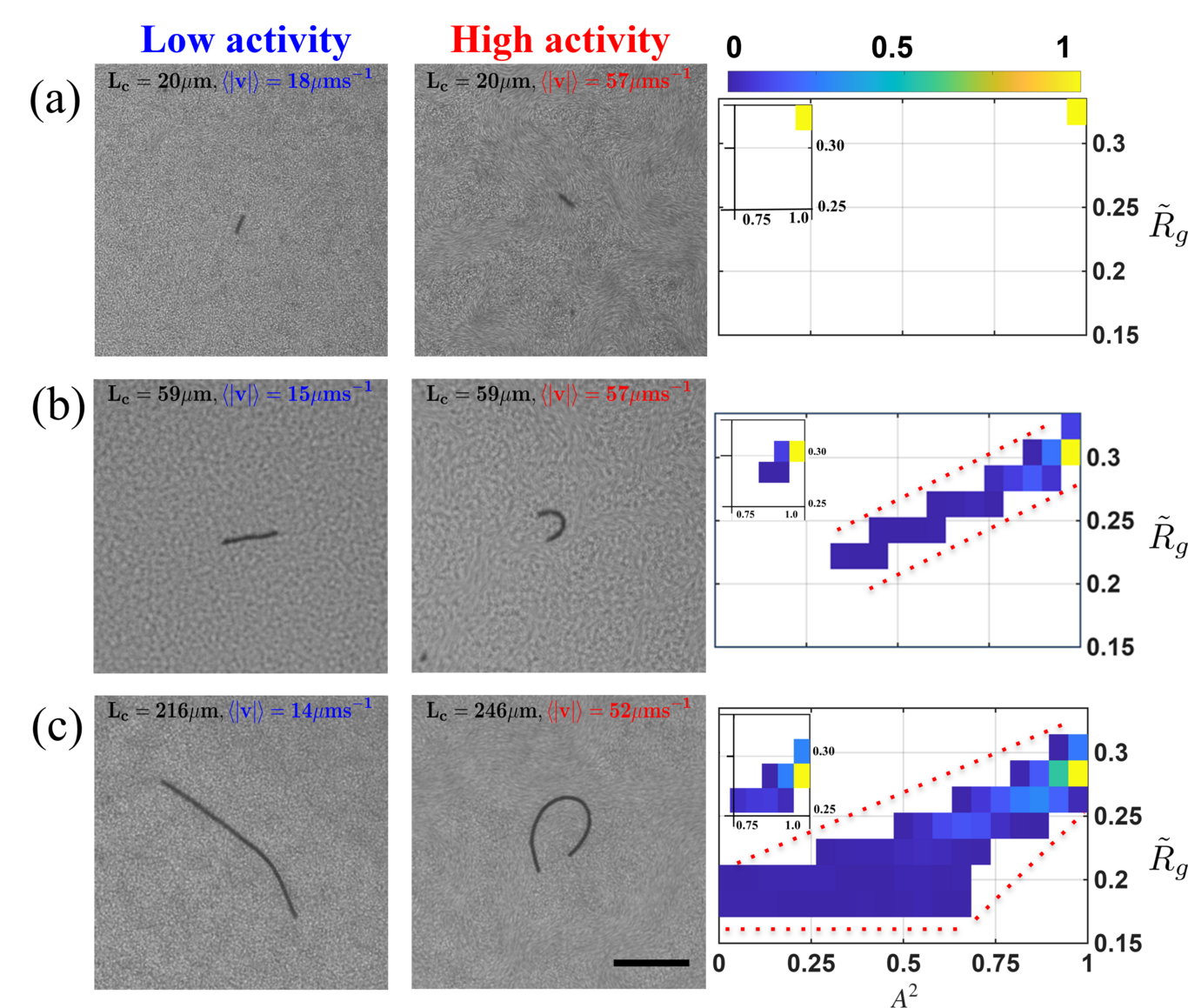}
\caption{{Conformations sampled by filaments of different contour lengths in
bacterial baths. (a) Short filaments remain straight in both low and
high activity baths, and occupy the corner of the~\(A^2-\tilde{R}_g\)
configuration space. (b) Intermediate filaments are bent into C-shapes
at high bath activity, and show an extended coverage in the
configuration space. (c)Long filaments show stronger deformations and a
richer variety of conformation states, represented by the expansion
of~\(A^2\) values at~low~\(\tilde{R}_g\) values. Color
bar: probability of conformation states normalized by the maximum
probability for each case. Scale bar:~ \(100\mathrm{\mu m}\)
{\label{469950}}%
}}
\end{center}
\end{figure}\selectlanguage{english}
\begin{figure}[h!]
\begin{center}
\includegraphics[width=0.70\columnwidth]{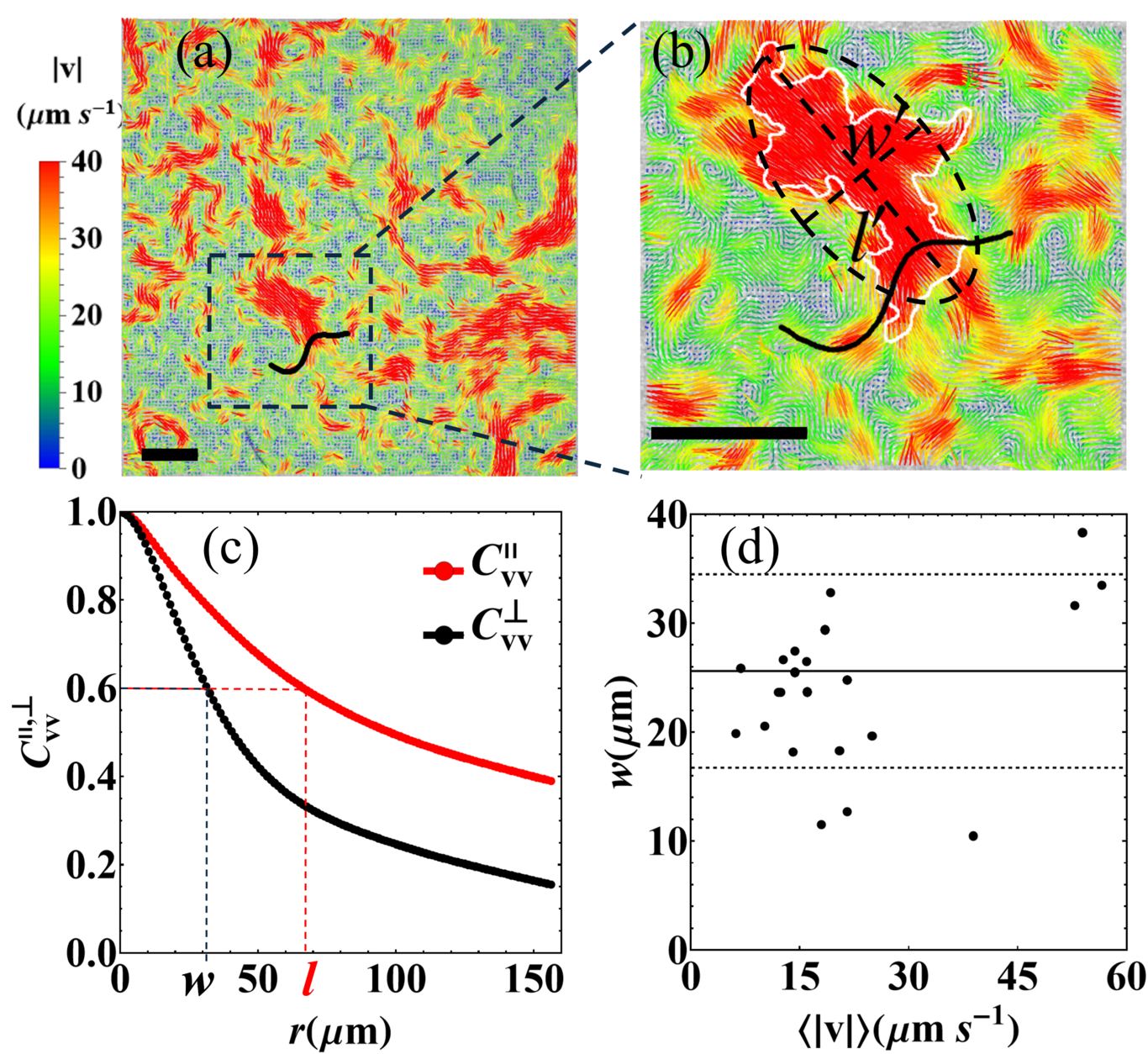}
\caption{{Characterization of the mesoscale flow in bacterial baths. (a) Direct
observation of jet-like structures impinging on and deforming filaments.
Representative image of the passive filament in the bath with color and
arrows corresponding to the local velocity field measured from PIV. (b)
Zoomed-in view showing the elongated shape of a typical jet marked by
the white boundary (SI).~{{Fitting the boundary with an ellipse results
in the length~}\(l^{\prime}\) and width~}\(w^{\prime}\) of the jet.(c) A typical anisotropic velocity correlation used to estimate
average~\(l\) and~\(w\) values for a given
active bath. (d) Jet widths~\(w\) for all active baths
studied, which shows no clear dependence on the activity. Scale
bar:~\(150\mathrm{\mu m}\).
{\label{786246}}%
}}
\end{center}
\end{figure}\selectlanguage{english}
\begin{figure}[h!]
\begin{center}
\includegraphics[width=0.70\columnwidth]{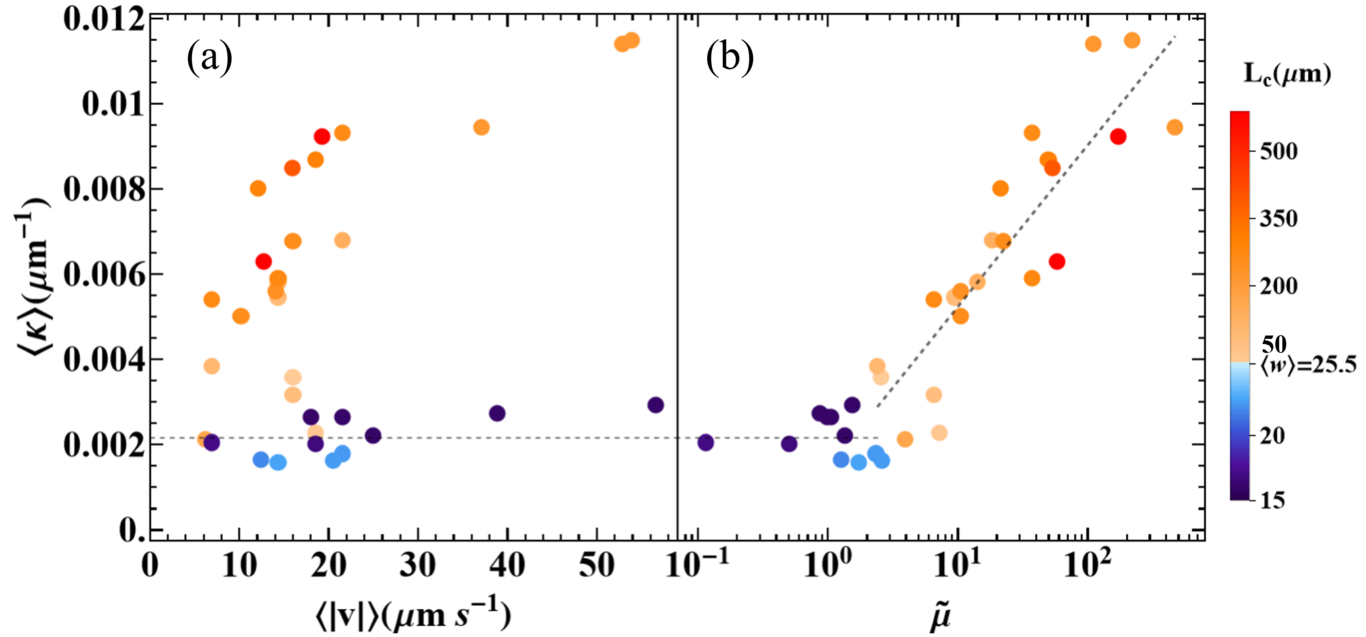}
\caption{{Competition between viscous drag and filament rigidity governs filament
configurations. (a) Average curvature~\(\langle \kappa \rangle\) of the
filaments at different activity levels of the bath show distinct
behaviors according to its length. Short filaments remain straight
regardless of bath activity level, while long filaments show increased
curvature as activity increases. (b) Average curvature shows a clear
threshold behavior on the dimensionless elastoviscous
number~\(\tilde{\mu}\), with a transition close to unity.~
{\label{174161}}%
}}
\end{center}
\end{figure}

\selectlanguage{english}
\FloatBarrier
\bibliographystyle{pnas2009}
\bibliography{biblio}
\end{document}